\documentclass[12pt,draftclsnofoot,onecolumn]{IEEEtran}
\usepackage{amssymb}
\usepackage{amsmath}
\usepackage{amsfonts}
\usepackage{graphicx}
\usepackage{subfigure}
\usepackage{cite}
\usepackage{enumerate}
\usepackage{url}
\usepackage[ruled]{algorithm2e}
\usepackage{algorithmicx}
\usepackage{mathtools}
\allowdisplaybreaks[4]

\begin{document}
\title{\huge{V2X Meets NOMA: Non-Orthogonal Multiple Access for 5G Enabled Vehicular Networks}}
\author{
\IEEEauthorblockN{
\normalsize{Boya Di}\IEEEauthorrefmark{1},
\normalsize{Lingyang Song}\IEEEauthorrefmark{1},
\normalsize{Yonghui Li}\IEEEauthorrefmark{2},
\normalsize{and Zhu Han}\IEEEauthorrefmark{3}\\}
\IEEEauthorblockA{\small
\IEEEauthorrefmark{1}School of Electrical Engineering and Computer Science, Peking University, Beijing, China, \\Email: $\left\{\mbox{diboya, lingyang.song} \right\}$@pku.edu.cn.  \\
\IEEEauthorrefmark{2}School of Electrical and Information Engineering, University of Sydney, Sydney, Australia, \\Email: yonghui.li@sydney.edu.au. \\
\IEEEauthorrefmark{3}Electrical and Computer Engineering Department, University of Houston, Houston, TX, USA, \\Email: hanzhu22@gmail.com.} \\

}

\maketitle
%\vspace{-0.2cm}
%Different from the traditional ad-hoc communications operated in unlicensed spectrum, D2D broadcast in licensed spectrum can be assisted in part by the network, just like cellular communications.
\begin{abstract}
%车载重要，LTE的优势，NOMA引入的优势。由于LLHR的限制，我们介绍了一些挑战和相关solution，我们propose了NOMA-HCHD，extended to 广播，我们也介绍了其他extensions.
Benefited from the widely deployed infrastructure, the LTE network has recently been considered as a promising candidate to support the vehicle-to-everything (V2X) services. However, with a massive number of devices accessing the V2X network in the future, the conventional OFDM-based LTE network faces the congestion issues due to its low efficiency of orthogonal access, resulting in significant access delay and posing a great challenge especially to safety-critical applications. The non-orthogonal multiple access (NOMA) technique has been well recognized as an effective solution for the future 5G cellular networks to provide broadband communications and massive connectivity. In this article, we investigate the applicability of NOMA in supporting cellular V2X services to achieve low latency and high reliability. Starting with a basic V2X unicast system, a novel NOMA-based scheme is proposed to tackle the technical hurdles in designing high spectral efficient scheduling and resource allocation schemes in the ultra dense topology. We then extend it to a more general V2X broadcasting system. Other NOMA-based extended V2X applications and some open issues are also discussed.
\end{abstract}
%%%%%%%%%%%%%%%%%%%%%%%%%%%%%%%%%%%%%%%%%%%%%%%
\section{Introduction}%
%%%%%%%%%%%%%%%%%%%%%%%%%%%%%%%%%%%%%%%%%%%%%%%
In the past few years, there have been greatly increasing applications of vehicular networks developed for future intelligent transportation, such as advanced driver assistance on active safety and traffic efficiency. To support various applications, an integrated system of the vehicular networking, namely vehicle-to-everything (V2X), has been proposed to enable vehicles to communicate with each other and beyond~\cite{SHPPZF-2016}. It provides three types of communications, i.e.,  vehicle-to-vehicle (V2V), vehicle-to-pedestrian (V2P), and vehicle-to-infrastructure/network (V2I/N) referring to the communication between a vehicle and a roadside unit/network. Although IEEE 802.11p has established security and upper layer specifications for V2X services, its unpredictable latency and limited transport capacity have confined its ability to achieve low latency and high reliability (LLHR). Recently, the widely deployed Long Term Evolution (LTE) networks~\cite{SS-2010} are being considered as a very promising solution for supporting V2X services~\cite{ACCIM-2015}, because large cell coverage, controllable latency, and high data rates can be achieved even in a high-mobility scenario.

%Paragraph1: development of vehicular network (from the prospective of traffic $\&$ application demand), V2X = V2V+V2I/N+V2P; from 11p to LTE-based V2X
Various techniques have been proposed in LTE to provide reliable communications, which can potentially be used to achieve LLHR in V2X applications. For V2I services naturally supported by LTE via the downlink/uplink transmission, the multimedia broadcast/multicast service (MBMS) can be utilized for the base station (BS) broadcasting~\cite{3GPP-2016} to achieve resource-efficient transmission, thereby reducing the latency to a relatively low level. For V2V services supported by the LTE device-to-device (D2D) communications~\cite{LDZE-2015}, end-to-end latency can be largely reduced via the direct link between users in proximity bypassing the BS, enabling the key support especially for the safety-critical applications.

Several works have discussed the proposal design for LTE-based V2X services to approach the LLHR requirement~\cite{SHPPZF-2016,WPL-2016,LJMMD-2016}. In~\cite{SHPPZF-2016}, enhanced design aspects to support LTE-based V2X services have been presented and a new demodulation reference signal sequence design has been performed. In~\cite{WPL-2016}, a distributed media access control (MAC) scheme has been proposed for the D2D OFDMA-based cellular networks. Fair sharing of spectrum among transmitter nodes has been performed to ensure the reliability of delivery. In~\cite{LJMMD-2016}, a unified radio frame structure and a MAC protocol have been proposed to enable reliable heterogeneous V2X services in different use cases.

However, as an orthogonal multiple access based (OMA-based) system, LTE mainly supports mobile devices sharing resources in an orthogonal manner, leading to serious congestion problems due to the limited bandwidth~\cite{ACCIM-2015}. To be more specific, resource collision may occur between OMA-based vehicle users, and the user access rate is difficult to be guaranteed via the orthogonal spectrum management in a dense moving environment. In a general V2X broadcasting scenario, this may also result in even more severe collision problems with significant packet loss~\cite{WPL-2016}, especially for a dense topology. Therefore, a more spectrally efficient radio access technology is still required for V2X services.

%Paragraph2: existing techniques in LTE-based V2X; remained problems
To tackle the challenges of access collision reduction and massive connectivity, non-orthogonal multiple access (NOMA) schemes have been introduced as a potential solution, which allow users to access the channel non-orthogonally by either power-domain~\cite{SKBNLH-2013} or code-domain~\cite{huawei-wp} multiplexing. Multiple users with different types of traffic requests can transmit concurrently on the same channel to improve spectrum efficiency and alleviate the congestion of data traffic, thereby reducing the latency. To make the NOMA scheme more practical, various multi-user detection (MUD) techniques, such as successive interference cancellation (SIC), are applied at the end-user receivers for decoding to cope with the co-channel interference caused by spectrum sharing among various users. In addition, uplink contention based NOMA~\cite{AZNYBVMZ-2014} has been proposed to reduce the control signaling overhead, especially for the small packet transmission, in order to overcome the shortage of large latency produced in LTE uplink.
%network-assisted 和 network-controlled没有说出来
%Paragraph3: motivation of NOMA (brings massive connectivity, and helps reduce resource collision).

Capable of achieving high overloading transmission given limited resources, NOMA provides a new dimension for V2X services to alleviate the resource collision, thereby improving the spectrum efficiency and reducing the latency. Due to the non-orthogonal nature of NOMA as well as the mobility and dense topology of vehicular network, the design of scheduling and resource allocation schemes for NOMA-based cellular V2X services becomes different from the traditional OMA-based system in many aspects, including:
\begin{itemize}
\item \textbf{Scheduling scheme:} In the OMA-based V2X network, semi-persistent scheduling (SPS) is applied in which the resources are booked by the vehicles every few transmission periods~\cite{3GPP-2016}. However, since the NOMA-based Rx decoding requires real-time channel state information (CSI) of multiple Tx users for enhancing the quality of SIC decoding, a new scheduling scheme combing the dynamic power control with SPS needs to be considered.
\item \textbf{Spectrum management:} Compared to the OMA-based case, the NOMA-based V2X network introduces co-channel interference by allowing multiple vehicles to share the same subband, which provides an extra dimension influencing the spectrum efficiency as well as user fairness.
\item \textbf{Power control:} Due to the dense topology of vehicular network, when the Tx users broadcast to their neighborhood, cross interference is brought to those Rx users in the overlapping region of multiple Tx users' communication range, which requires the design of new power control strategy of each Tx user in the NOMA-based V2X network.
\item \textbf{Signaling control:} Due to the requirement of prior knowledge for joint decoding such as the CSI of Tx users, signaling between the Rx users and Tx users for information exchange is an important issue which couples with the power control of Tx users. In the traditional OMA-based case, such prior knowledge is usually provided by the BS, which may introduce great latency to V2X applications.
\end{itemize}

In this article, we discuss the applicability of the NOMA technique for supporting V2X services starting with the basic V2X unicast systems and then extending to other typical scenarios as listed below.
\begin{itemize}
\item \textbf{NOMA-based V2X unicast systems:} The basic V2X unicast model consists of multiple V2V pairs sharing the same channel simultaneously for direct communication. One receiver (Rx user) may suffer from co-channel interference from neighboring transmitters (Tx users), and one Tx user may cause interference to multiple Rx users nearby. SPS-based spectrum management of the BS and dynamic distributed power control of the users need to be considered given the requirement of LLHR.
\item \textbf{NOMA-based V2X broadcast systems:} V2X broadcasting is essential for safety-critical applications, in which each vehicle is required to broadcast a small data packet containing safety-critical information to its neighborhood within every short transmission period. Beside of the issues brought up in the unicast case, time domain resource allocation and Tx-Rx selection of the BS need to be further considered for interference management.
\item \textbf{NOMA-based uplink V2I networks:} The uplink V2I networks consist of multiple devices transmitting to the network/infrastructure by utilizing the contention-based code-domain NOMA technique~\cite{AZNYBVMZ-2014}. Main issues include resource allocation based codebook selection and contention-based backoff mechanism design.
\item \textbf{NOMA-based V2V networks with multiple operators:} In such a network, at lease a set of carriers for direct communication is shared by the vehicles subscribed to different operators, and joint data transmission and reception is performed by the operators to improve the spectrum efficiency of cell-edge vehicle pairs. Coordination between the operators to jointly allocating antennas, spectrum, and power is required.
\end{itemize}
% 新的资源分配问题产生了，具体有四个场景，每个的问题简单描述。

The rest of this article is organized as follows. Section~\uppercase\expandafter{\romannumeral2} provides an overview of existing cellular V2X techniques and NOMA-based communications. A NOMA-based mixed centralized/distributed (NOMA-MCD) scheme is proposed in Section~\uppercase\expandafter{\romannumeral3} for resource allocation and signaling control in the aforementioned basic V2X unicast system. Main challenges, possible solutions, and performance evaluations are discussed in detail. Other NOMA-based V2X communication extensions are presented in Section~\uppercase\expandafter{\romannumeral4} focusing on both benefits and research problems brought by NOMA. In the final section, we draw the main conclusions, and also discuss some open problems and potential research directions.

\section{Overview of Cellular V2X and NOMA-based communications}
\subsection{LTE-supported V2X Services}
Traditionally, V2X services can be classified as two types, safety-critical and traffic efficient applications, with different requirements of packet size and latency. Safety-critical messages are usually short broadcast messages with strict latency constraints, putting high requirement on scheduling and packet loss. Different from the safety-critical messages, traffic efficient messages refer to a large amount of sensed data of vehicles sent to the infrastructure or other vehicles. Instead of achieving stringently low latency, the key point of traffic efficient message transmission is how to achieve continuous communication in a moving environment~\cite{WSZJ-2013} while guaranteeing the delivery of other human-to-human (H2H) traffic.

\begin{figure}[!t]
\centering
\includegraphics[width=3.7in]{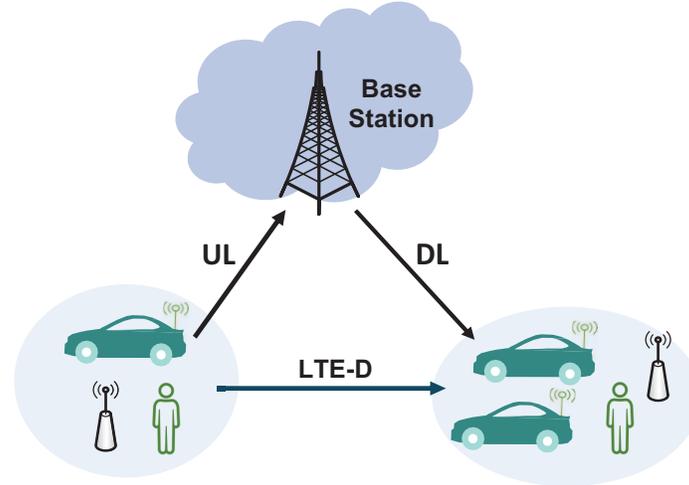}
\caption{A basic framework illustrating two communication modes in LTE-based vehicular network.} \label{framework}
\end{figure}

To support both safety-critical and traffic efficient applications, LTE has provided two communication modes for V2V and V2I, i.e., LTE-direct (LTE-D) and cellular UL/DL, respectively, as illustrated in Fig.~\ref{framework}.
\begin{itemize}
\item \textbf{LTE-D for V2V:} LTE-D refers to direct communication between two devices bypassing the BS in a proximal way. The design of LTE-D has made up for the native feature of LTE that message passing between vehicles via the infrastructure may produce large end-to-end latency. Resource allocation problems involving multiple D2D pairs and cellular users have been discussed in the literature~\cite{XSHZWCJ-2013}. However, unlike traditional H2H LTE-D networks, the V2V network has brought new concerns such as intractable cross interference due to the moving environment and a dense topology.

%LTE本来不支持终端直通，但是D2D的出现使车之间的通信成为可能，减少了延迟。D2D中资源碰撞避免主要通过正交的user association来解决。

\item \textbf{Cellular UL/DL for V2I:} Cellular UL/DL refers to the common communication mode between devices and the BS/road side units. For the case in which the density of vehicles is particularly sparse, cellular UL/DL can be considered for assistance. Note that multiple transmissions via unicast in the DL may lead to great resource waste given the expensive licensed spectrum. To achieve resource-efficient transmission, BS broadcasting/multicasting can be adopted via MBMS. Based on coordination between multiple cells, MBMS helps improve the cell-edge performance and reduce the latency.
%对于车辆不怎么密集的情况，可以基站辅助通信，为了resource efficient，采用广播；广播的好处是提升边缘用户体验，减少latency
\end{itemize}

Note that the mobility of vehicles leads to the rapid variation of fading, which poses a great challenge to the scheduling schemes for V2X services. Various existing scheduling schemes in LTE are listed as below suitable for different latency requirements:
\begin{itemize}
\item In \textbf{Dynamic Scheduling}, users are allocated resources for every transmitted packet in each time slot based on real-time CSI, which requires accurate channel estimation. It is suitable for sudden and frequently size-varying data transmissions that potentially consume wide bandwidth.
\item In \textbf{Semi-Persistent Scheduling}, the BS allocates the predefined sets of resources to those users requesting for transmission every SPS period. The length of each SPS period is set as the same order of the required latency. By removing unnecessary signaling exchange in each slot, SPS reduces a great deal of latency, which is particularly suitable for transmitting periodically short messages with a fixed packet size, such as the basic safety messages.
\end{itemize}
%既适用于PC5，也适用于Uu

\subsection{Non-Orthogonal Multiple Access Technique}
NOMA has been proposed as a new access technique for next generation mobile communications, supporting massive connectivity and sufficient spectrum usage. Two types of NOMA schemes have drawn great attention as comprehensively introduced below:
\begin{itemize}
\item \textbf{Power domain NOMA (PD-NOMA)} allows multiple users to share the same channel simultaneously by power domain multiplexing at the Tx, and SIC can be applied at the end-user Rx users to decode the received signals which suffer from co-channel interference. It smartly exploits the differences of received power levels to obtain higher spectrum efficiency than the OMA scheme. Industry standards have been widely discussed for future deployment~\cite{docomo-wp}, and several efficient algorithms for the resource allocation have been proposed~\cite{DSL-2016}.
\item \textbf{Code domain NOMA (CD-NOMA)} is also known as SCMA, short for sparse code multiple access. It uses sparse (or low-correlation) spreading sequences to integrate data streams of various users and then spread over multiple subchannels to realize overloading. Each user is identified by a codebook containing multiple codewords, and one codeword is represented by the spreading sequence of which length equals to the size of subcarrier set. At the transmitter, bit streams of each user are directly mapped to different sparse codewords of the corresponding codebook. All mapped codewords are then multiplexed over the dedicated subchannels, followed by a near-optimal detection of over-laid receiving sequences benefited from the sparsity of codewords. Protocols for the SCMA schemes have been released in~\cite{huawei-wp}.
\end{itemize}

\section{NOMA Applicability to Cellular V2X} %NOMA怎么用到LTE-V2X中
In this section, we elaborate how to apply the PD-NOMA technique in the cellular vehicular network for resource collision reduction, thereby achieving low latency. To better explain this, we first present a basic V2X unicast model in which the cross interference is considered. The NOMA-MCD scheme is then proposed and evaluated as below. Extension of this scheme to a more general V2X broadcast case will be presented in detail in Section~\uppercase\expandafter{\romannumeral4}.A.

\begin{figure}[!t]
\centering
\includegraphics[width=5.0in]{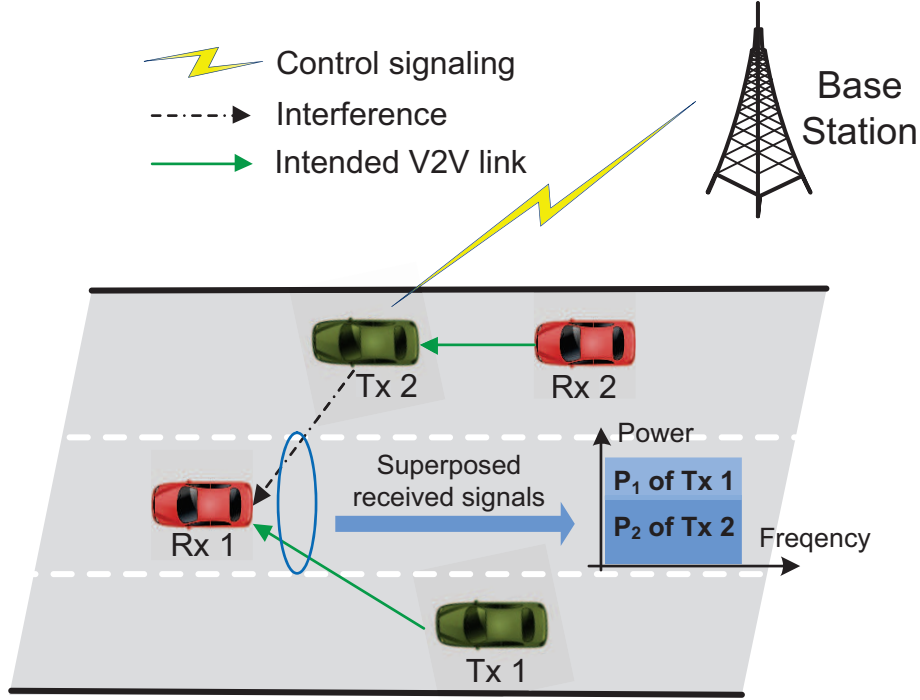}
\caption{System model of the NOMA-based V2X unicast systems.} \label{system_model1}
\end{figure}

\subsection{NOMA-based V2X Unicast System Model}
Consider a NOMA-based V2X unicast network as shown in Fig.~\ref{system_model1}. Multiple V2X pairs communicate in a NOMA-based mode in which one sub-channel can be assigned to multiple pairs, and one V2X pair can occupy multiple sub-channels. The BS is capable of frequency resource allocation requiring global position information of the network. For those conflicting Rx users who suffer co-channel interference from neighboring Tx users, SIC is performed based on the decreasing channel gains of Tx users for decoding the target signals. To improve the system performance, we aim at maximizing the total number of successfully decoded packets, for which the data rate of a target signal exceeds a given threshold. Due to the severe cross interference caused by overlapping interference ranges of each Tx user as well as the prior knowledge requirement of SIC decoding, new challenges have emerged in the resource allocation and signaling control as presented in sequel.

\subsection{Key Problems and Solutions of Resource Allocation and Signaling Control}
%先说这两个的重要性，再说完全的dynamic和完全的半静态都有问题;
Before presenting the key problems, we first illustrate the reason why we adopt a mixed centralized/distributed scheduling scheme as follows. Considering the mobility of vehicles and complicated cross interference caused by the dense topology, selection of scheduling scheme is a non-ignorable factor affecting both latency and data rates. Traditional dynamic centralized resource allocation may cause significantly large delay since the users need to send resource request messages to the BS for every data packet. In addition, accurate CSI is hard for the BS to obtain in a mobile environment. For a dynamic distributed scheduling scheme, CSI can be updated in time via direct links. It may be suitable for the case where the BS is out of function; however, it is very costly and unwise for the users to access the channel in a contention-based method within the coverage of the BS. Full centralized SPS has a good performance in reducing the delay; nevertheless, it fails to capture the rapidly changing CSI caused by the mobility, leading to potentially large resource collision during each SPS period.
%The expected added value of network control

To achieve the optimal scheduling given the latency requirement and mobility features, we adopt a mixed centralized/distributed scheme in which the BS performs the SPS and the Tx users perform distributed autonomous power control in each time slot. At the beginning of each SPS period, the BS determines how to allocate frequency resources to the Tx users, which takes full advantage of the global position information obtained by the BS to perform the interference management. Dynamic distributed power control by the Tx users is then performed to resolve the issues that the BS cannot obtain real-time CSI as well as to improve the rate performance of PD-NOMA.

Below we discuss the main problems and possible solutions in detail from the perspectives of centralized spectrum management of the BS and distributed power control.
\subsubsection{Centralized Spectrum Management of the BS}
%给出条件和问题，简要描述matching算法;  %重传，中断传输等机制并没有讨论
To reduce the resource collision, the BS performs frequency resource allocation based on the position information of each vehicle updated at the beginning of each SPS period. Different from the traditional OMA-based case, co-channel interference needs to be considered here.

Due to mobility of vehicles, not only the full CSI is very costly for the BS to acquire, but also the CSI can get easily outdated due to the rapid variation of small scale fading. Therefore, we adopt partial CSI containing the path loss and shadowing during each SPS period. Define an indicator variable $x_{j,k}$ to denote whether sub-channel $k$ is allocated to Tx $j$. Considering the reliability of the network, we aim at improving the number of successfully decoded signals, which can be approximated by the sum of continuous logistic functions\footnote{Here we take the logistic function as an example to depict the successful decoding probability. Other approximation methods such as the experience-based SINR-packet reception ratio curve can also be used.} as shown below:
\begin{equation} \label{subchannel_allocation}
\{ {x_{j,k}}\}  = \mathop {\arg }\limits_{{x_{j,k}}} \max \sum\limits_{j,k} {\prod\limits_{\scriptstyle j' \in {{\cal{S}}_j} \cup \left\{ j \right\} \hfill \atop \scriptstyle {x_{j',k}} = 1 \hfill} {\frac{1}{{1 + {e^{ - \eta \left( {{Rat}{e_{j',k}} - Rat{e_{th}}} \right)}}}}} },
\end{equation}
in which ${Rate}_{th}$ denotes the minimum data rate required for successful decoding, $Rat{e_{j',k}}$ denotes the data rate of the link Tx user $j'$ -- Rx user $k$, ${\cal{S}}_j$ represents the set of Tx users with higher channel gains~\footnote{According to the SIC principle, the signal of Tx user $j$ can be successfully decoded by Rx user $k$ if the signal of Tx user $j'$ with a higher channel gain can be decoded by the Rx user first.} than Tx user $j$ over subchannel $k$, and $\eta$ is the slope parameter of the logistic function.

Note that this is a non-convex problem due to the binary variables, which can be converted into a many-to-many matching problem with externalities. Considering the Tx users and sub-channels as two sets of objects to be matched, the BS performs a swap-matching algorithm briefly described as below. Initially each Tx user randomly selects a set of sub-channels based on its priority. In the following swap-matching phase, the BS keeps searching for two pairs of Tx and sub-channel to check whether they can swap their matches such that the total utility in $\left( \ref{subchannel_allocation} \right)$ can be improved. If these two pairs exist, the BS swaps the original matches of them. The largest number of one player's matches is fixed during the matching phase, and the iteration stops until there exists no blocking pair in current matching.

\subsubsection{Distributed Power Control of the Users}
%\subsubsection{Control information}

After the frequency resource allocation at the beginning of each SPS period, dynamic power control is then performed by the Tx users in each time slot. The transmit power of each Tx user has cross influence over the set of neighboring Rx users, posing a great challenge to perform the interference management in a distributed manner, especially for the PD-NOMA scheme. Specifically, though a Tx user intends to communicate with only one target Rx, there may exist multiple Rx users within this Tx user's interference disk, each of which affected by the transmit power of this Tx user. Similarly, the target Rx user may also receive from multiple Tx users, rendering the decoding effect unstable. In addition, we note that necessary prior knowledge is required by the Rx users to perform the SIC decoding, such as the number of Tx users in the interference range and the corresponding CSI. Therefore, the distributed power control requires information exchange between the Tx users and the Rx users, i.e., the control signaling, as will be explained below.

 %Therefore, a specific MAC control structure needs to be designed such that the potential signaling cost is limited to a tolerable level.

We divide each transmission slot into one control signaling portion and one data transmitting portion, in which the control signaling portion consists of multiple Tx-Rx iterations for control message exchange between the Tx users and the Rx users. Power control strategy of each Tx user is determined in every iteration. To limit the potential signaling costs to a tolerable level, we assume that there are $T_c$ Tx-Rx iterations in the control portion, followed by the data transmission. Each iteration consists of one Tx block and one Rx block, and works as below. In the Tx block, every Tx user adjusts its transmit power so that the transmitted reference signals can be successfully decoded by its target Rx user while causing minimum interference to other Rx users. Each Rx user obtains its neighboring Tx users' CSI and transmit power via the received reference signals. In the following Rx block, the Rx users then calculate the potential co-channel interference brought by Tx users in the neighborhood, and send back to corresponding Tx users for further processing in the next Tx block.

In each Tx block, every Tx user adjusts its transmit power based on the feedback sent by the Rx users. To avoid the situation where each Tx user transmits with the maximum power, the power control strategy for each Tx user is set as below: if the co-channel interference caused by a Tx user is larger than a threshold, its transmit power is set to zero; otherwise, the power is set as the minimum value such that the rate of the direct link between this Tx user and its target Rx user is larger than the decoding rate threshold. Small scale fading is considered during the power control.

%is obtained by
%\begin{subequations}\label{power_control}
%\begin{align}
%{p_j} &= \mathop {\arg }\limits_{{p_j}} \min {\rm{Rat}}{{\rm{e}}_{j,m}} \\
%\textbf{\emph{s.t.:}} & {\rm{Rat}}{{\rm{e}}_{{\rm{j,k}}}} \ge {\rm{Rat}}{{\rm{e}}_{{\rm{th}}}},\ p_j \le {p_{th}},
%\end{align}
%\end{subequations}
%in which ${\rm{Rat}}{{\rm{e}}_{j,m}}$ represents the data rate of the direct link between Tx $j$ and Rx $m$, and the small scale fading is considered.

\subsection{Performance Evaluation}
In summary, our proposed NOMA-MCD scheme is described as below. At the beginning of each SPS period, each vehicle user updates its position and velocity information to the BS. The BS then allocates the frequency resources to the Tx users to maximize the number of successfully transmitted messages in which large scale fading is considered based on predictable distance information. After the centralized spectrum assignment, distributed power control coupled with the Tx-Rx selection is then performed. In each iteration of the control signaling portion, the Tx users adjust their transmit power based on the feedback from neighboring Rx users. The whole distributed power control process ends within the control portion of a transmission slot, followed by the data transmission in which Rx users can decode received signals given the CSI obtained from the control portion.

\begin{figure}
\centering
\subfigure[]{
\label{alphatwo:a} %% label for first subfigure
\includegraphics[width=3.2in]{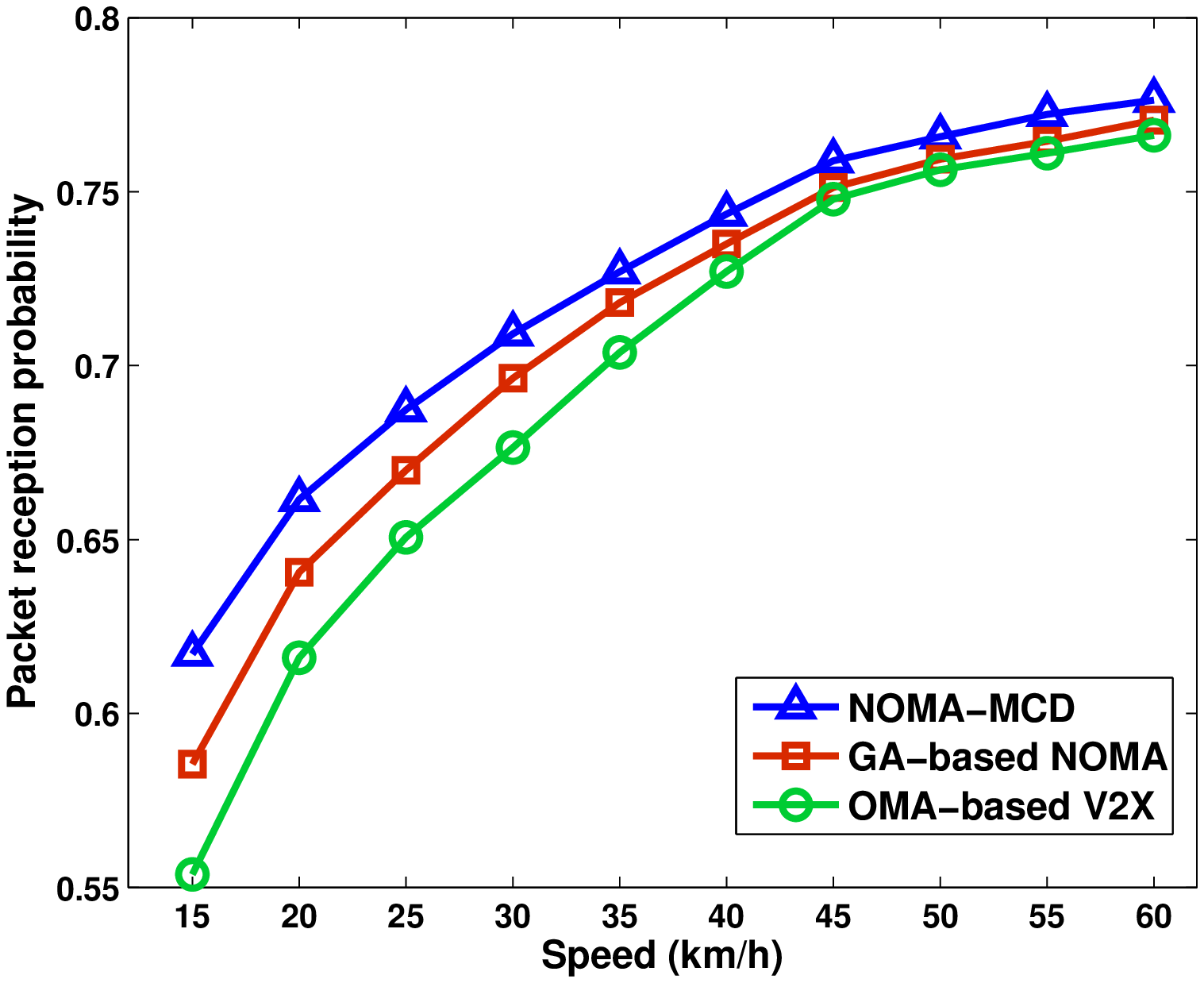}}
\hspace{-0.3in}
\subfigure[]{
\label{alphatwo:b} %% label for second subfigure
\includegraphics[width=3.2in]{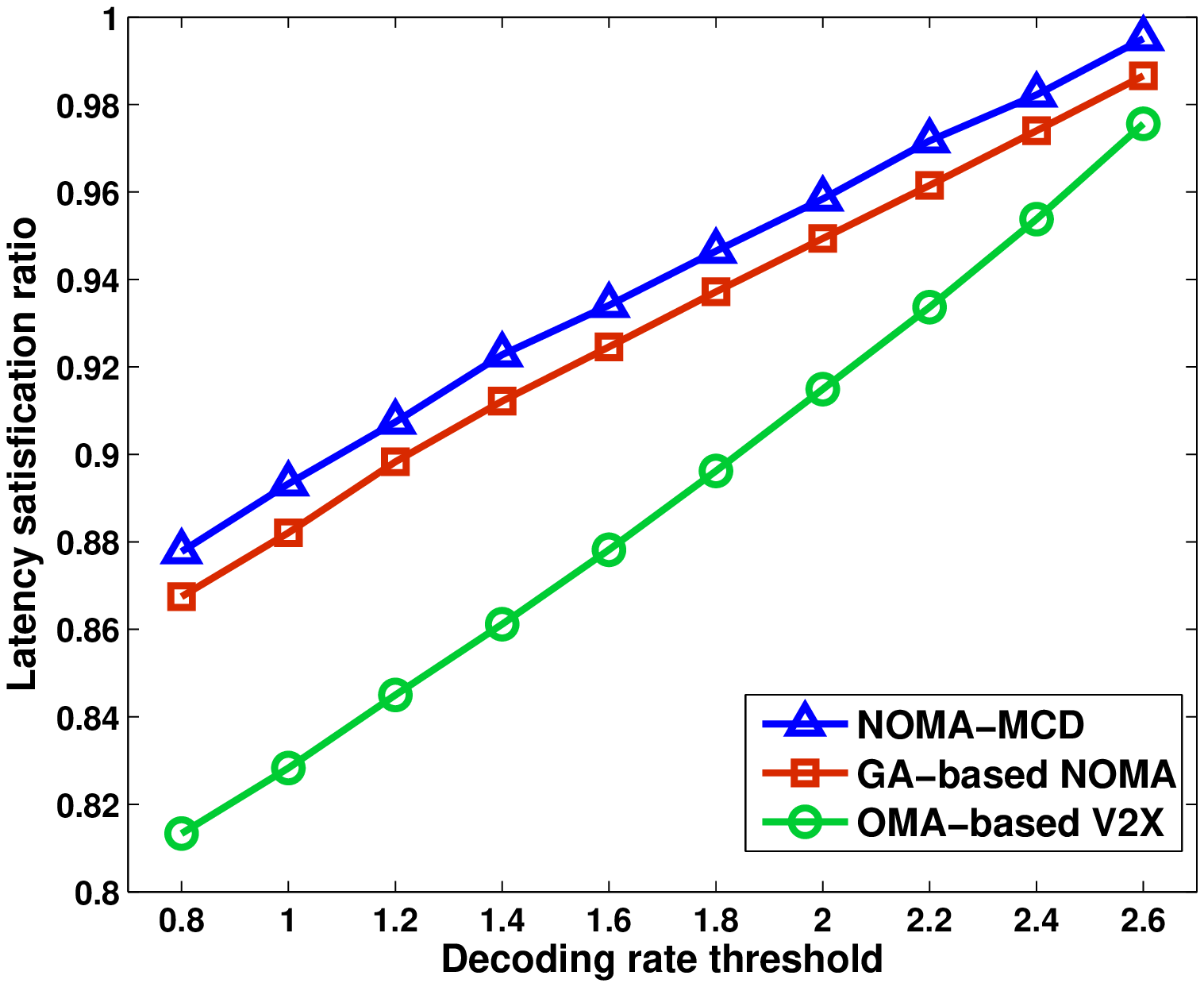}}
\caption{System performance of the NOMA-MCD scheme.} \label{simulation} %% label for entire figure
\end{figure}

To evaluate the performance of NOMA-MCD scheme, we compare it with the traditional OMA-based LTE-D scheme with respect to the packet reception probability as well as the latency performance\footnote{Latency satisfaction ratio refers to the ratio between the number of successfully transmitted signals (given the latency constraint) and that of successfully decoded signals.} as shown in Fig.~\ref{simulation}. We assume that 20$\%$ and 80$\%$ vehicles on the road serve as Tx users and Rx users, respectively, and at most 2 Tx users can share the same subchannel. It is observed that the NOMA-MCD scheme performs better than the OMA-based scheme.

%SPS period, length of control portion, weighted value of each user
%说一下解调的时候信令的事

\section{NOMA-based V2X Communication Extension}
In this section, we extend the basic V2X model introduced in Section~\uppercase\expandafter{\romannumeral3} to other applications in the cellular vehicular network. Some key research problems as well as possible solutions are also discussed.

%\begin{table}[!h]
%\begin{center}
%\caption{NOMA-based V2X Communication Extension}
%\label{table_2}
%\begin{tabular}{|p{24mm}|p{20mm}|p{45mm}|p{45mm}|}
%\hline
%\bf{NOMA-based V2X application} & \bf{Category} & \bf{Intended traffic demand} & \bf{Key challenges}\\
%\hline NOMA-based V2X broadcasting systems & Mixed centralized/distributed & Each vehicle broadcasts to its neighborhood at least once every transmission period & Cross interference and hidden terminal problems\\
%\hline NOMA-based uplink V2I networks & Distributed & Each vehicle uploads to the infrastructure & High latency caused by LTE UL\\
%\hline NOMA-based V2V Networks with Multiple Operators & Mixed centralized/distributed & Multiple cell-edge V2V pairs subscribed to different operators perform direct communications & Joint resource allocation of multiple operators\\
%\hline
%\end{tabular}
%\end{center}
%\end{table}

\subsection{NOMA-based V2X Broadcast Systems}
\begin{figure}[!t]
\centering
\includegraphics[width=5.4in]{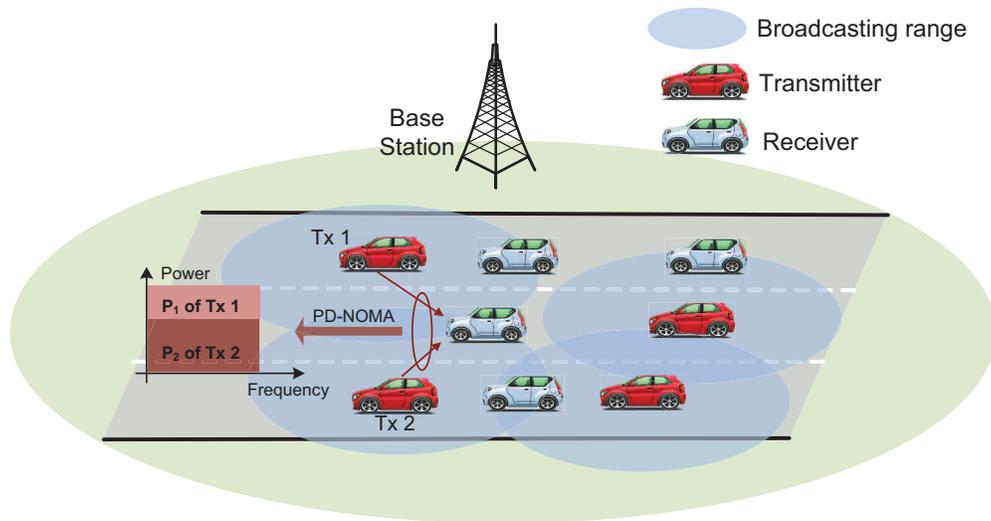}
\caption{System model of the NOMA-based V2X broadcast transmission.} \label{system_model2}
\end{figure}

%enhancing LTE with support for D2D broadcast is viewed as a key enabler for the use of LTE technology. as the basis for future public safety communication networks
%描述场景需求，传统的OMA怎么做->为什么引入NOMA；另起一段说明HCHD可以用，但是要考虑时间维和用户选择，比单播更复杂一下，会变成一个三维matching问题；此外对于power control，考虑给用户分优先级
V2X broadcasting via the direct links is viewed as a key support for utilizing the LTE technology in vehicular networks, which enables safety-critical applications. Take a dense V2X broadcast transmission scenario for example, as shown in Fig.~\ref{system_model2}. In every transmission period consisting of multiple time slots, each vehicle is required to broadcast a small data packet containing safety-critical information at least once to its neighborhood. In the traditional OMA-based LTE-D mode, hidden terminal problems exist in which a conflicting Rx user may lie in the overlapping section of multiple Tx users' communication ranges, and these Tx users are not close enough to communicate with each other. Existing solutions for this problem usually focus on the collision avoidance mechanism design in which only one Tx user can transmit in current slot while other Tx users in the neighborhood yield to this Tx user and keep silent. However, this may result in insufficient use of the spectrum resources and loss of time-validity for the sate-critical messages of those silent Tx users. % 需要通过避让来解决干扰问题

The proposed NOMA-MCD scheme is then naturally extended to this case such that one Rx user can receive from multiple Tx users simultaneously, reducing the resource collision and improving massive connectivity via power-domain multiplexing. Different from the V2X unicast version, time-domain resource management and user scheduling need to be considered, and the power control strategy of each Tx user is re-considered, as illustrated below.
\subsubsection{Tx-Rx selection and time-frequency resource allocation of the BS}
During each time slot, a user can only be either the Tx user or the Rx user due to the half-duplex nature, and thus, any two users in each other's communication range cannot be assigned to transmit simultaneously in one SPS period. At the beginning of each SPS period, the BS decides which two subsets of vehicles act as Tx users and Rx users in each time slot, respectively, and how time-frequency resources are allocated to the Tx users. Therefore, the centralized resource allocation problem for the BS is then formulated as a three-dimensional integer programming problem.

\subsubsection{Power control strategy of each Tx user}
Unlike the Tx users in the unicast case each of which has only one target Rx user while other neighboring Rx users are interfered, Tx users in the broadcast case aim at adjusting the transmit power so as to maximize the number of successfully decoding Rx users in the neighborhood. However, it is almost impossible for every Rx user to successfully decode all the superposed received data in a dense environment. Therefore, each Tx user minimizes its transmit power during the process of control signaling such that a certain percent of neighboring Rx users in the communication range are guaranteed to successfully decode the received signals.

\subsection{NOMA-based Uplink V2I Networks}
%解决了上行latency大的问题，而且提供了massive connectivity
Note that the LTE uplink transmission can be a bottleneck of achieving LLHR due to the excessive signaling exchange between each vehicle and the BS. Specifically, this issue can degrade the performance of safety-critical applications in which short packets containing basic safety information are frequently updated by the vehicles. Aiming at finding a low-latency access technique, we observe that the contention based SCMA scheme may be a suitable candidate.

\begin{figure}[!t]
\centering
\includegraphics[width=6.2in]{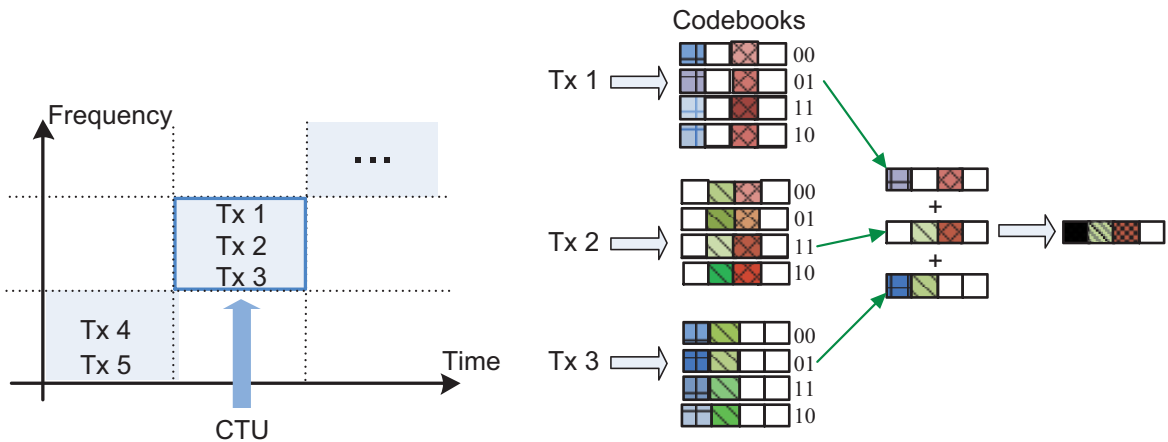}
\caption{NOMA-based uplink V2I networks: codebook-based resource occupation, encoding, and multiplexing} \label{system_model4}
\end{figure}

%先阐述SCMA怎么工作，再给出可以做的部分(结合了802.11p和NOMA的优点）
A SCMA-based uplink V2I network works as in Fig.~\ref{system_model4}, where each vehicle transmits to the BS by competitively occupying one or more contention transmission units (CTU) from the dedicated contention region. Given the contention region which is part of the UL bandwidth, one CTU refers to a combination of time, frequency, SCMA codebook, and pilot sequence. At the beginning of each SPS period, the CTUs are determined for the users either by the BS assigning or implicitly deriving from the user ID. When multiple users are assigned the same CTU for data transmission, user collision occurs, which can be resolved through the random back-off mechanism similar to that in 802.11p. No transmission grant of the BS is required before the users send packets, which is different from the traditional centralized UL transmission. For receiver decoding, the BS attempts to detect received packets by utilizing all possible assigned access code sequences, in which the message passing algorithm (MPA) for joint decoding can be utilized.

 %903 页右侧最上一段；再说一下解调 % granted free

Enabling the grant-free UL transmission while allowing system overload via multiplexing, the SCMA scheme has provided a new solution to achieve massive connectivity and meet very stringent latency requirements for some V2X services. Note that the number of vehicles set in one subband influences the collision probability of CTUs, thereby having an impact on the decoding performance of MPA. Therefore, a trade-off between the massive connectivity and latency needs to be discussed especially in an ultra dense network. In addition, resource allocation based codebook selection and back off mechanism design for SCMA are also crucial factors influencing the reliability of data services as well as latency performance.

\subsection{NOMA-based V2V Networks with Multiple Operators}
\begin{figure}[!t]
\centering
\includegraphics[width=5.4in]{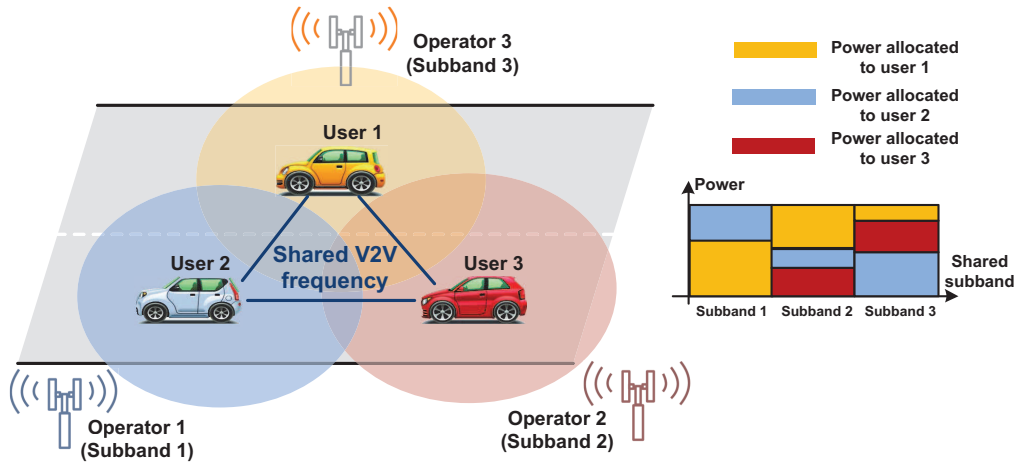}
\caption{NOMA-based V2X networks with multiple operators.} \label{system_model3}
\end{figure}

%为了解决干扰问题，两个BS本来就需要相互通信协调；那么为了提升频谱效率，还不如顺便用NOMA，反正都协作了
Consider a V2V network with multiple operators as shown in Fig.~\ref{system_model3}. A set of carriers for direct communication is shared by the vehicles subscribed to different operators, i.e., vehicles belonging to different operators may transmit on the same carrier~\cite{3GPP-2016}. For the cell-edge vehicles, cooperation between multiple operators is necessary for joint spectrum management and power control to achieve reliable direct communication. More than one cell-edge V2V pairs may contest for the shared frequency resources, leading to resource collision.

To improve the spectrum efficiency of cell-edge users, NOMA-based joint data transmission and reception of the vehicles can be performed assisted by the operators. Due to the mobility of vehicles, dynamic cell hand-off needs to be considered in the proposed scheme. The association between users and operators should be carefully designed to obtain the prior knowledge of joint decoding while maintaining low latency of the system. In addition, precoding of the BS may be different from the traditional single-cell case, since it is not easy for a precoder associated with multiple separate antennas to form the physical beam that perfectly fits the distribution area of those co-channel NOMA users.

%The cell-edge vehicles may experience interference from neighboring cells. To mitigate the inter-cell interference, cooperation between multiple operators is necessary In traditional OMA-based system, conflicting operators coordinate to assign orthogonal sub-channels to these cell-edge users via the spectrum occupation negotiating, rendering the frequency resources not fully utilized.

%Another method is joint decoding of the cell-edge users' signals across neighboring cells, which requires multiple operators to obtain the knowledge of both the users' data and CSI. Large amount of extra information needs to be transmitted to a cell-edge user for decoding including other users' data in the neighboring cell.

%预编码需要考虑；移动性增加了这个网络的复杂性。
%\begin{figure}[!t]
%\centering
%\includegraphics[width=4.4in]{system_model3.eps}
%\caption{Networking NOMA assisted LTE-V2N/V Network with Multiple Operators.} \label{system_model3}
%\end{figure}

%总结优势：codebook reuse;overloading; the number of vehicles set in one subband influences the collision probability of CTUs, thereby influencing the decoder performance of the Message Passing Algorithm, i.e., tradeoff between reliability and latency.

%\subsection{Cognitive NOMA}

\section{Conclusions and Future Outlook}
In this article, we introduce the NOMA technique into the LTE-based vehicular network to support massive connectivity and reduce resource collision for multiple V2X applications via either the power-domain or code-domain multiplexing. The new scheduling scheme, resource allocation algorithm, and control structure are designed in which the expected added value of network control is exploited to meet the requirement of LLHR. The proposed NOMA-MCD scheme is illustrated in detail given a basic V2X unicast model, and simulation results indicate that it can efficiently reduce the resource collision compared to the traditional OMA-based scheme. An explicit extension of this scheme to a more general safety-critical V2X broadcast scenario is then elaborated. Other NOMA-based extended V2X applications are also presented such as networking NOMA and contention based SCMA, which can be applied to cope with the multiple-operator case and LTE-uplink latency, respectively.

As an effective approach supporting massive connectivity and high spectrum efficiency, the NOMA technique shows its potential to enhance the quality of cellular V2X services. Several open issues still need to be carefully addressed before practical implementation of the NOMA-based cellular vehicular network, which may drive the future inventions and research. These issues include synchronization in a high-mobility case, co-existence of LTE and WiFi, and coordination between various types of services/devices with different requirements. Two future research problems to be discussed in this field are listed below as examples.
\begin{itemize}
\item \textbf{NOMA for cooperative V2X:} Consider a broadcast scenario in which the originating node (BS or vehicles) broadcasts superposed signals to multiple Rx users. Lack of feedback from the Rx users to the originating node makes the retransmission not user-specific. Note that in the NOMA scheme, users with good channel gains are capable of decoding the information intended to other users with poor channel gains. Such prior information can be utilized for cooperation-based retransmission, in which a Rx user with good channel gain forwards the messages to other Rx users with poor channel gains via a direct link~\cite{DPP-2015}. Considering the mobility of vehicles, research topics such as dynamic spectrum management and communication protocol design need to be redeveloped.
\item \textbf{NOMA for cognitive V2X:} For traffic efficient applications, a large amount of data generated by the vehicles poses great pressure on traditional H2H traffics. To guarantee the demand of H2H traffics while improving the quality of data transmission of vehicles, a new scheme named cognitive NOMA can be utilized in which the vehicles opportunistically access the channels which are originally occupied by the cellular users. Given a dedicated spectrum band, cellular users and vehicles are regarded as primary users and secondary users, respectively. Vehicle users can only access the channel when the services of cellular users are not affected. Different from the traditional cognitive radio scheme, power control needs to be performed carefully and a scheme coordinating both primary and secondary users should be designed.
\end{itemize}
%latency; mobility; synchronization; heterogeneous network

\end{document}